\documentclass[10pt,doublesided, doublecolumn,final]{IEEEtran}

\usepackage{float}
\usepackage{algorithm}
\usepackage[noend]{algorithmic}
\usepackage[geometry]{ifsym}
\usepackage {ifsym}
\usepackage{bbm}
\usepackage{amssymb}
\usepackage{yhmath}
\usepackage{amscd}
\usepackage{dsfont}
\usepackage{amsmath}
\usepackage{epsfig}
\usepackage{graphics}
\usepackage{psfrag}
\usepackage{rotating}
\usepackage{amsmath,amsthm,}
\usepackage{amsfonts}
\usepackage{url}
\usepackage{color}
\usepackage{float}
\usepackage{balance} 
\usepackage{mathtools}
\usepackage{tabularx}
\usepackage{booktabs}
\usepackage{enumitem} 
\usepackage{caption}
\usepackage[justification=centering]{caption}
\usepackage[labelfont=bf,format=plain,justification=raggedright,singlelinecheck=false]{caption}

\newtheorem{theorem}{Theorem}

\newtheorem{proposition}{Proposition}

\newtheorem{corollary}{Corollary}
\newtheorem{remark}{Remark}

\newcommand{\bs}{\boldsymbol}

\newcommand{\defn}{\stackrel{\triangle}{=} }

\renewcommand{\d}{\mathrm{d}}

  \makeatletter

\begin{document}
\title{Information Energy Capacity Region for SWIPT Systems over Raylegh-Fading Channels   }

 \author{Nizar Khalfet, \IEEEmembership{Member,~IEEE,} and Ioannis Krikidis, \IEEEmembership{Fellow,~IEEE}
\thanks{N. Khalfet and I. Krikidis are with Department of Electrical and Computer Engineering, University of Cyprus, Cyprus (e-mail: \{khalfet.nizar,~krikidis\}@ucy.ac.cy).}
}
\maketitle

\begin{abstract}
In this paper, we study the fundamental limits of simultaneous information and power transfer  over a Rayleigh-fading channel, where the channel input is constrained to  peak-power (PP) constraints that vary in each channel use by taking into account  high-power amplifier (HPA) nonlinearities.  In particular, a three-party communication system is considered,  where a  transmitter aims simultaneously conveying information to an information receiver  and delivering energy to an energy harvesting  receiver. For the special case of static PP constraints,  we study the information-energy capacity region and the associated input distribution under: a) average-power and PP constraints at the transmitter, b) an HPA nonlinearity at the transmitter, and c) nonlinearity of the energy harvesting circuit at the energy receiver. By extending Smith's mathematical framework \cite{Smith-1971},  we show that the optimal input distribution under those constraints is discrete with a finite number of mass points. We show that HPA significantly reduces the information energy capacity region. In addition, we derive a closed-form expression of the capacity-achieving distribution for  the low PP regime, where there is no trade-off between information and energy transfer. For the case with time-varying PP constraints, we prove that the optimal input distribution has a finite support by using Shannon's coding scheme. Specifically, we numerically study a particular scenario for the time-varing PP constraints, where the PP constraint probabilistically  is either zero or equal to a non-zero constant.
\end{abstract}



\section{introduction}
Simultaneous information and power transfer (SWIPT) is a technology that exploits the duality of the radio frequency (RF) signals, which can carry both information and energy \cite{Bi15} through appropriate co-design and engineering. The idea of wireless power transfer (WPT)  was first proposed by Tesla in the 20-th century \cite{Tesla-Patent-1914}, and now presents a promising solution for modern communication systems such as  low-power short-range communication systems, sensor networks, machine-type networks, and body-area networks \cite{KRI1}. The  notion of the information-energy capacity region for SWIPT systems, it was first formalized by Varshnay \cite{Var} in the context of point-to-point scenarios. This work has been extended in \cite{ShannTesla} for a parallel links point-to-point channel. More recent works study the integration of SWIPT  to more complex network topologies e.g.,  multiple access channel \cite{BPKP-TIT-2017}, interference channel \cite{KP-JSAC-2019}, multiple-input multiple-output \cite{PC-WC-2013}, multiple-antenna cellular networks \cite{DiRenzo}, etc. A comprehensive overview of existing results in SWIPT for various fundamental multi-user channels is presented in \cite{Belhadj-Amor-ICT-2016}.\\
The design of the WPT component  is crucial in order to characterize SWIPT systems.
Most of the literature assumes  simple linear models for the RF energy harvester (EH) receiver \cite{PC-WC-2013, Zhang-TWC-2013} to simplify analysis. However,  one of the main particularities of a SWIPT network is that the WPT channel is highly nonlinear (in contrast to the linear information transfer channel). Recent studies take into account the nonlinearity of the rectification circuit, and study the impact of waveform design and/or input distribution on the achieved information energy capacity region. For instance, the work in \cite{BClerckx2016} models the rectifier behavior and introduces a mathematical framework to design waveforms that exploit nonlinearity.  This observation introduces a relevant question for SWIPT networks: "what is the fundamental limits of a SWIPT system with a non-linear EH receiver ?" \\
The problem was first formalized in \cite{Bruno-Arkv-2018} by considering a truncated Taylor expansion series approximation for the diode's characterization function over an additive white Gaussian noise (AWGN) channel. The authors have shown that the optimal input distribution is zero-mean complex Gaussian distribution, with an asymmetric power allocation for the real and the imaginary parts. However, a more general model was proposed in \cite{Rania-2018}, by using the exact form of the diode's characteristic function. The authors have extended Smith's mathematical framework \cite{Smith-1971} and have shown that the optimal input distribution under the first the and second moments statistics as well as a peak power (PP) constraint at the transmitter, it is unique, discrete with a finite number of mass points.\\
 On the other hand, experimental studies demonstrate that signals with high peak-to-average-power-ratio (PAPR) e.g., multi-sine, chaotic signals, white noise,  etc \cite{6766272}, provide a  higher direct-current (DC) output, in comparison to constant-envelop sinusoidal signals  \cite{ 8476597, 7580639}. However, signals with high PAPR are more sensitive to high-power amplifier (HPA) nonlinearities, which  significantly degrade the quality of the communication \cite{Aissa-2010, 9170573}.  With the exception of a few studies (e.g., \cite{9174172}), existing works do not consider the effects of HPA on SWIPT performance and assume that the HPA operates always in the linear regime. In addition, most of the aforementioned studies on SWIPT systems focus on simple AWGN channels and therefore the impact of the channel fading has not been investigated. To the best of the authors' knowledge, this is the first work that takes into account the effect of HPA's non-linearity  on SWIPT systems over fading channels from an information theoretic standpoint. \\
 
 In practice, energy is harvested from nature through different sources, such that solar, vibrations, wind, etc. Therefore, the energy available at the transmitters is not always a deterministic process and can be modeled as a random process that varies in time \cite{6216430}. In such systems, the transmitter casually observes the arrived energy and then makes a decision on the code symbol transmitted. This setup is modeled as a state-dependent channel with causal state information available at the transmitter.  The fundamental limits of the state-dependent channel was first introduced by Shannon in \cite{Shann}, where a capacity achieving coding scheme was proposed.  By extending the framework of  \cite{Smith-1971} and  \cite{Shann}, the authors characterized the optimal input distribution over an AWGN channel with time-varying PP constraints by extending the alphabet of the codewords in accordance with the PP constraints. The later work was extended  for a Gaussian multiple access channel in \cite{MAC}, where  the authors have shown that the boundary of the capacity region is achieved by a discrete input distribution with a finite support. However, none of these works addresses the fundamental limits of the fading channel with time-varying PP constraints in a SWIPT context.\\

This paper studies the fundamental limits of SWIPT over a Rayleigh-fading channel, where the channel input is constrained to a PP varying constraint at each channel use; in addition, we take into account a memoryless HPA model at the transmitter and a non linear power transfer channel. The varying PP constraint induces a state-dependent channel with a perfect causal state information at the transmitter, since the amplitude process is observed by the transmitter.  In particular, we consider a basic three-node SWIPT system,  where a transmitter simultaneously sends data to an information receiver and power to an EH receiver through a Rayleigh-fading channel; we consider both average power (AP) and time varying PP constraints at the transmitter as well as a non-linear power transfer channel. For the special case of  static PP constraints, we characterize the information energy capacity region and we show that the associated capacity-achieving input distribution  is unique, discrete,  with a finite number of mass points. For the general case with time varying PP constraints,  we prove that the capacity achieving distribution has a finite support by using Shannon's coding scheme \cite{Shann}. Our study generalizes the result for the capacity-achieving input distribution of a conventional discrete-memoryless Rayleigh-fading channel under AP, which has been studied in \cite{Shemai-2001}. We show that HPA significantly reduces the information energy capacity region, while increasing the PP constraint enlarges the associated region. Finally, we  study the optimal input distribution for the low PP regime, where a no trade-off between information and energy transfer is observed.

The technical contribution of this paper is twofold: 

\begin{itemize}

    \item For scenarios with static PP constraints at the transmitter (e.g., the transmitter is connected to the power grid), we show that the optimal input distribution that maximizes the information energy capacity is unique, discrete, with a finite number of mass points. We numerically study the impact of the HPA non-linearity on the information energy capacity region and we show that HPA significantly reduces the corresponding region. In addition,  we propose a mathematical framework to study the optimal input distribution for the low PP regime, where there is not a  trade-off between information and energy transfer.
   
    \item For  scenarios with time-varying PP constraints at the transmitter (e.g., the energy arrives at the transmitter in a sporadic way according to a random process),  we characterize the information energy capacity region by extending the results in \cite{Var}. Firstly, we construct a channel which has an input extended by the cardinality of the state alphabet  and we optimize the input distribution  by applying Shannon's strategy \cite{Shann}. Then, we show that the capacity achieving distribution for the extended channel has a finite support. For the sake of exposition, we deal with a scenario of a binary on-off energy arrival process, where either a constant  amount of energy  or no energy arrives.
\end{itemize}



\begin{table}[t]
\centering
\begin{tabular}{ |p{1cm}||p{2.5cm}||p{1cm}|p{2.5cm}|  }
\hline
\textbf{Notations} &\textbf{Description}&\textbf{Notations} &\textbf{Description}\\
 \hline
   $n$   &  Number of channel uses & $X$ & Random variable of the input signal\\
    \hline
   $M$ &  Cardinality of the states &  $\bs{X}$  & Random variable of the extended alphabet\\
    \hline
  $t$  & Time index & $X^{\star}$  & Optimal input random variable \\
     \hline
 $P$   &  Average power constraint    &$F^{(n)}$&   Sequence of the the input distribution\\
 \hline
$E_{\text{req}}$&  Energy required at the EH &  $F(\cdot)$ &Probability distribution function of $X$\\
 \hline
 $h_{1,t}$ & Channel fading for the information link& $I(\cdot)$&   Mutual information as a function of the distribution $F$\\
 \hline
$h_{2}$   &Channel fading for the EH link & $F^{\star}$&  Optimal input probability distribution \\
 \hline
 $F^{(n)}$ &  Sequence of a distribution $F$  &  $i(x;F)$&mutual information density of $F$ evaluated at $x$\\
 \hline
 $F_N^{\star}$& Optimal  distribution with $N$ mass points  & \hspace{1ex}$\mathcal{CN}\hspace{-0.5ex}(0,\hspace{-0.5ex} \sigma_1^2)$  &Circular complex Gaussian random variable with zero mean and $\sigma_1^2$ variance\\
 \hline
  $\mathrm{Supp}(F)$ & Support of a distribution $F$& $E_0$&  Point of increase of  $F^{\star}$ \\
 \hline
\end{tabular}
\caption{Summary of notation.}
\end{table}

 \emph{Notation:} In this paper, sets are denoted with uppercase calligraphic letters. Random variables are denoted by uppercase letters, e.g., $X$. The realization and the set of the events from which the random variable $X$ takes values are  denoted by $x$ and $\mathcal{X}$, respectively. The argument $\mathds{E}[X]$ denotes the expectation with respect to the distribution of a random variable ${X}$. The notation $F$ denotes the probability distribution function of a random variable $X$, and $F^{\star}$ represents the optimal input distribution. The notation $\mathrm{Supp}(F)$ is the support of a distribution $F$, i.e,  $\mathrm{Supp}(F)=\{x \in \mathcal{X} | F(x) \neq 0\}$. Table. I summarizes the key notation of the paper. \\
 The reminder of this paper is structured as follows. The system model is presented in Section \ref{SecSystemModel}. In Section \ref{SecStatic}, we characterize the information energy capacity region for the case of  static PP constraints. For the time-varying PP constraint case, we study the optimal input distribution in Section \ref{Sectime}. Finally, numerical results are presented in Section \ref{SecNum}.

\section{System model}
\label{SecSystemModel}
 Consider a three part communication system, where a transmitter aims simultaneously convey information to an information receiver (IR) and power to an EH receiver. The IR converts the received signal to the baseband to decode the transmit information, while the EH receiver harvests energy from the received RF signal. In each channel use, the transmitter inputs a pulse-amplitude modulated signal $x(t)=\sum_{k=-\infty}^{\infty}x[k]p(t-kT)$, with an average power $P$,  where $p(t)$ is the rectangular pulse shaping filter (i.e., $p(t)=1$ for $0<t\leq T$), $T$ is the symbol interval, and $x[k]$ is the information symbol at time index $k$, modeled as the realization of an independent and identically distributed (i.i.d) real random variable $X$ with cumulative distribution function (CDF) $F$. We assume a normalized symbol interval $T=1$ and thus the measures of energy and power become identical and therefore are used equivalently. The system model is depicted in Fig. 1. The transmitted amplitude-modulated signal $x(t)$ is subjected to nonlinearities induced by the HPA;  the output of the nonlinear HPA can be written as $\hat{x}[k]=d(x[k])$ (i.e., random variable $\hat{X}=d(X)$), where $d(\cdot)$ denotes the AM-to-AM conversion which is given by the considered solid state power amplifier (SSPA)  HPA model \cite{Aissa-2010} i.e., 
\begin{align}
&d(r)=
\frac{r}{\left[1+\left(\frac{r}{A_{s}} \right)^{2\beta} \right]^{\frac{1}{2\beta}}}, 
\end{align}
where $A_{s}$ is the output saturation voltage, and $\beta$  represents the smoothness of the transition from the linear regime to the saturation. 

\begin{figure}[t]
\label{FigSystem}

\centering
\includegraphics[width=1\linewidth]{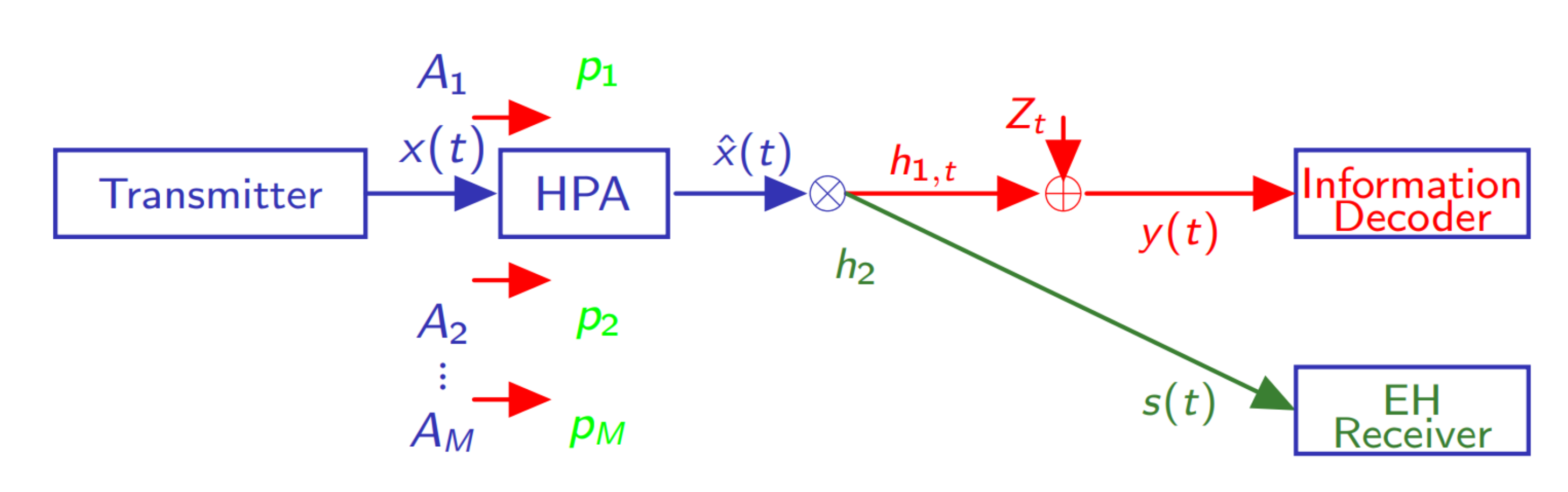}
\caption{A SWIPT system over a fading channel with time varying PP constraints, HPA at the transmitter, and a non linear EH channel.}
\end{figure}

\subsection{Information transfer}
For the information transmission, we consider a memoryless discrete-time Rayleigh-fading channel \cite{Shemai-2001}; the channel output at the receiver during the channel use $t$ is given by  
\begin{equation}
    y(t)=h_{1,t} \hat{x}(t) + Z_t,
\end{equation}
where $\hat{x}(t)$ is the channel input induced by the HPA, $y(t)$ is the channel output, and $h_{1,t}$ and $Z_t$ are independent complex circular Gaussian random variables  distributed as  
\begin{IEEEeqnarray}{cCl}
h_{1,t} &\sim& \mathcal{CN}(0, \sigma_1^2),\\
Z_t &\sim& \mathcal{CN}(0, \sigma_2^2).
\end{IEEEeqnarray}
Since the phase of the fading parameter $h_{1,t}$ is uniform, an equivalent channel with a nonnegative input $\hat{X}$ and a nonnegative output  $Y$ is proposed in \cite{Shemai-2001} and is given by

\begin{equation}
p(y|\hat{x})=\frac{1}{1+\hat{x}^2}\exp\left(-\frac{y}{1+\hat{x}^2}\right).
\end{equation}
\subsection{Power transfer}
At the EH receiver, the contribution of the noise is assumed to be negligible, hence, the representation of the received signal at the EH receiver (in the baseband equivalent) is given by
\begin{equation}
    s(t) = h_{2}\hat{x}(t),
\end{equation}
where $h_2 \in \mathds{R}$ is the channel fading for the link between the transmitter and the  EH receiver; it is assumed to be constant/static  in this work\footnote{In practical SWIPT environments, the EH receivers are located close to the transmitter (e.g., a line of sight link). Hence, we can assume that the channel fading $h_2$ (associated to the power transfer) is a constant \cite{Rania-2018}.}. Let  $\mathcal{E} : \mathds{R} \rightarrow \mathds{R}_+$ be the function that determines the average energy harvested \cite{Rania-2018}, which is given by
\begin{equation}
    \mathcal{E}(\hat{X})=\mathds{E}\left[I_0\left(\sqrt{2}Bh_2|\hat{X}|\right)\right], 
\end{equation}
		 where $I_0(\cdot)$ is the modified Bessel function of the first kind and order zero and $B$ is a constant that depends on the characteristics of the rectification circuit.  The EH constraint is reduced as
		 \begin{equation}
		  \mathcal{E}(\hat{X})\geq E_{\mathrm{req}}.
		 \end{equation}
\subsection{Problem formulation}

 The transmitter is subjected to a time-varying PP constraint as shown in Fig. 1. Let $A$ be a PP constraint random variable with alphabet $\mathcal{A}=\{a_1,a_2,\ldots,a_M\}$, where $M$ is the cardinality of the alphabet of the energy arrivals.  Let also $\{A_k\}_{k=1}^{\infty} \in \mathcal{A}$ be the i.i.d. PP constraint process with 
 \begin{equation}
    \mathrm{Pr}\left[A_k=a_i\right]=p_i, \hspace{2ex} \forall \hspace{1ex} k,
\end{equation}
with $i \in \{1,2,\ldots, M\}$. The realization of the PP constraint $\{A_1,A_2,\ldots,A_M\}$ is observed by the transmitter, and the code symbols must satisfy
\begin{equation}
    |X_k| \leq A_k, \hspace{2ex} k \in \{1,2,\ldots,n\},
\end{equation}
where $n$ is the number of channel uses. 

Note that the variation of the amplitude at the transmitter is not known at the receiver; in this case the channel is   state dependent  with a causal state information available only at the transmitter. By applying the capacity scheme proposed by Shannon \cite{Shann}, we define a random variable $\bs{X}=[X_1,X_2,\ldots, X_M]$ that takes values into the set $\mathcal{X}= [0,A_1] \times [0,A_2] \ldots [-A_M,A_M] $, then the space of joint probability distribution function over $\mathcal{X}$ is given by  
			\begin{IEEEeqnarray}{lCl}
			\nonumber
			    \mathcal{F}_{A_1,A_2,\ldots,A_M}&\defn& \bigg\{ F : \int_{0}^{A_1} \int_{0}^{A_2} \ldots  \int_{0}^{A_M} \d F(x_1,x_2,\ldots,x_M)\\
			    &&=1 \bigg\}.
			\end{IEEEeqnarray}
The expression of the equivalent channel  is given by 
\begin{IEEEeqnarray}{l}
\nonumber
 f(y|\hat{x}_1,\hat{x}_2,\ldots, \hat{x}_M)=\sum_{i=1}^Mp_ip(y|\Hat{x}_i)\\
  \label{eqChannel}
 =\sum_{i=1}^Mp_i\frac{1}{1+\hat{x_i}^2}\exp\left(-\frac{y}{1+\hat{x_i}^2}\right).
			\end{IEEEeqnarray}
In particular, the mutual information density for the equivalent channel  \eqref{eqChannel} is given by
\begin{IEEEeqnarray}{l}
\nonumber
    i(x_1,x_2,\ldots,x_M;F)= \int f(y|x_1,x_2,\ldots,x_M)\\
    \times \log \left(\frac{f(y|x_1,x_2,\ldots,x_M)}{f(y;F)}\right) \d y.
\end{IEEEeqnarray}

By using Shannon strategy, the information energy capacity region is obtained by maximizing the mutual information between $\bs{X}$ and $Y$. The optimal input distribution that achieves the information energy capacity region for the time-varying PP constraint fading channel is obtained by solving the following optimization problem

			\begin{equation}
			\label{EqOptVArOr}
\begin{aligned}
& \underset{F \in \mathcal{F}_{A_1,A_2,\ldots,A_M}}{\text{sup}}
& & I(F)  \\
& \text{subject to}
& & \sum_{i=1}^M p_i\mathds{E}[X_i^2]\leq P, \\
& & & \sum_{i=1}^M p_i \mathcal{E}(\hat{X}_i) \geq E_{\mathrm{req}},
\end{aligned}
\end{equation}
where 
\begin{IEEEeqnarray}{r}
\nonumber
    I(F)= \int_{0}^{A_1} \int_{0}^{A_2} \ldots \int_{0}^{A_M}   i(x_1,x_2,\ldots,x_M;F)\\
    \d F(x_1,x_2,\ldots,x_M).
\end{IEEEeqnarray}
\subsection{Smith's framework \cite{Smith-1971}}
The capacity of  PP-constrained Gaussian channel was first introduced by Smith in \cite{Smith-1971}, where it was shown that the optimal input distribution that achieves the information capacity is discrete, unique,  with a finite number of mass points. The basic steps of the framework are summarized as follows: \begin{enumerate}
    \item \emph{Step 1}:  Prove that the capacity achieving distribution is unique by showing that the space of distribution functions is concave in a weak topology.
    \item \emph{Step 2}:  The dual problem is given by using the Lagrangian theorem \cite{Smith-1971}, since the basic  optimization is valuable for non constrained functions.
    \item \emph{Step 3}: Provide  a necessary and sufficient condition for the capacity achieving distribution  by using the optimization theorem.
     \item \emph{Step 4}: Prove that the optimal input distribution is discrete through contradiction, by using the identity theorem for  analytic complex functions.
\end{enumerate}


\section{SWIPT for  static PP constraint}
\label{SecStatic}
In this section, we focus on the special case  where the PP constraint at the transmitter is static/constant, i.e., $\mathcal{A}=\{a_1\}$. This case refers to SWIPT scenarios, where the transmitter has a constant power supply e.g., it is connected to the power grid.  Hence the formulation in \eqref{EqOptVArOr} is reduced to an one-dimensional optimization problem, and the input distribution is subjected to 
\begin{equation}
   |X_k|= |X| < A_1=A, \hspace{2ex} \forall \hspace{1ex} k.
\end{equation}
The conditional probability  in \eqref{eqChannel}  is reduced to 
	\begin{equation}
		p(y|\hat{x})=\frac{1}{1+\hat{x}^2}\exp\left(-\frac{y}{1+\hat{x}^2}\right).
	\end{equation}

The main objective is to maximize the average mutual information between $X$ and $Y$ subject to both AP and PP constraints on the transmit symbols $X$,  and a minimum harvested power constraint at the EH receiver. The optimization problem in \eqref{EqOptVArOr}  is reduced  as 
\begin{equation}
\label{EqOptimization}
\begin{aligned}
& \underset{\mathcal{F}_A \in \mathcal{F}}{\text{sup}}
& & I(F)=\int \int p(y|x)\log \frac{p(y|x)}{p(y; F)}\d y \d F(x), \\
& \text{subject to}
& &  \mathds{E}[X^2] \leq P, \\
& & &\mathcal{E}(\hat{X}) \geq E_{\mathrm{req}},
\end{aligned}
\end{equation}
where $\mathcal{F}_A$ is the set of all input distributions that satisfies the PP constraint, i.e., 
\begin{equation}
    \mathcal{F}_A=\bigg\{F \in \mathcal{F}; \int_{0}^{A} \d F(x)=1   \bigg\}.
\end{equation} 
 
The mutual information between $X$ and $Y$, as well the AP and PP constraints and the minimum harvested energy  required at the EH,  could be expressed in function of the  distribution $F$.
Denote by  $g_1: F \rightarrow \Re $ and $g_2: F \rightarrow \Re $ the following functions 
\begin{IEEEeqnarray}{cCl}
         I(F)&\defn&\int_{0}^{A}   i(x;F)   \d F(x),\\
         g_1(F)&\defn& \int_{0}^{A} x^2 \d F(x)-P,\\
         g_2(F)&\defn& E_{\mathrm{req}}- \int_{0}^{A} \mathcal{E}(\hat{x}) \d F(x). 
\end{IEEEeqnarray}
Denote by $\Omega$ the set of the constrained input distribution, such that
\begin{equation}
    \Omega=\bigg\{F \in \mathcal{F}; \int_{0}^{A} \d F(x)=1 ; g_i(F) \leq 0; i \in \{1,2\}  \bigg\},
\end{equation}
then the optimization problem in \eqref{EqOptimization} can be written simply as  $C= \sup_{F \in \Omega}I(F)$. 

\subsection{Discreteness of the optimal input distribution }
In this section, we study the properties of the capacity achieving distribution, i.e., the solution of the optimization problem in \eqref{EqOptimization}. By extending the mathematical framework  in \cite{Smith-1971}, Theorem \ref{TheoremUnique} establishes the existence and the uniqueness of the optimal input distribution (first step of Smith's framework). By using the Lagrangian Theorem, the dual equivalent problem is given by Corollary \ref{CoroLagrangien} (second step of Smith's framework). Furthermore, we give necessary and sufficient conditions for the optimal input distribution in Corollary \ref{CoroNecessary} (third step of Smith's framework). Finally, we show that the capacity-achieving input distribution is discrete in Theorem \ref{TheoremDiscrete}.

\begin{theorem}
\label{TheoremUnique}
The capacity $C$ is achieved by a unique input distribution $F^{\star}$, i.e,  
\begin{equation}
\label{eqprob}
C= \sup_{F \in \Omega}I(F)=I(F^{\star}).   
\end{equation}
\end{theorem}
	\begin{IEEEproof}
	The proof is presented in Appendix \ref{ProofOfTheoremunique}.
	\end{IEEEproof}	 
\begin{remark}
 Note that for the case where the EH constant is omitted, the problem is reduced to the conventional problem of  information capacity over a Rayleigh fading channel \cite{Shemai-2001}, where  the optimal input distribution is discrete with a finite number of mass points (even without considering the AP constraint). 
 \end{remark}

To establish a necessary and sufficient condition for the optimal input distribution, we refer to the optimization theorem \cite{Smith-1971}, which determines the optimal elements for a non-constrained convex space.  In the following Corollary, we transform the optimization problem in \eqref{eqprob}, to the dual problem so we are able to apply the basic optimization theorem presented in \cite{Smith-1971}.
\begin{corollary}
\label{CoroLagrangien}
    The strong duality holds for the optimization problem in \eqref{eqprob}, i.e.,  there exist constants $\lambda_1 \geq 0$ and $\lambda_2 \geq 0$ such that 
         \begin{equation}
             C=\sup_{F \in \mathcal{F}_A} I(F)-\lambda_1g_1(F)-\lambda_2g_2(F),
         \end{equation}
\end{corollary}
\begin{IEEEproof}
     The proof is presented in Appendix \ref{ProofOfCoroLagrangien}.
\end{IEEEproof}

The following Theorem  establishes a necessary and sufficient condition on the optimal input distribution.

\begin{theorem}
 \label{TheoremNecessary}
  $F^{\star}$ is the capacity achieving input distribution,  if and only if,  $\forall$  $F \in \Omega$, there exist $\lambda_1 \geq 0$ and $\lambda_2 \geq 0$ such that 
  \begin{equation}
  \label{eqnec}
          \int i(x;F^{\star}) -C-\lambda_1g_1(F)-\lambda_2g_2(F) \d F(x) \leq C-\lambda_1P+\lambda_2E_{\mathrm{req}}.
  \end{equation}
  \end{theorem}
 \begin{IEEEproof}
 The proof is presented in Appendix \ref{ProofOfNecessarry}.
 \end{IEEEproof}
By using the condition in \eqref{eqnec}, we derive a necessary and sufficient condition for the optimal input distribution in the following corollary. 
\begin{corollary}
\label{CoroNecessary}
Let $E_0$ be the points of increase\footnote{The points of increase of a distribution $F$ is the set of points, in which $F$ has a zero probability, i.e., $E=\{ x \in \mathcal{X}; F(x) \neq 0\}.$} of a distribution $F^{\star}$, then $F^{\star}$ is the optimal input distribution, if there exist $\lambda_1 \geq 0$ and $\lambda_2 \geq 0$, such that 
	\begin{IEEEeqnarray}{l}
\nonumber
	\lambda_1\left( x^2-P\right)-\lambda_2\left(I_0(\sqrt{2}Bh_2x)-E_{\mathrm{req}}\right)+C	\\
		\label{EqNESU}
	-\int p(y|x)\log \frac{p(y|x)}{p(y; F^{\star})}dy\geq 0,
	\end{IEEEeqnarray}
	for all $x$, with equality if $x \in E_0$.
\end{corollary}
%
\begin{IEEEproof}
The proof is presented in Appendix \ref{ProofOfCoroNecess}.
\end{IEEEproof}
By using the  invertible change of variables, i.e.,  $S=\frac{1}{1+X^2}$, we have 
	\begin{equation}
	p(y|s)=s\exp(-ys), \hspace{2ex} s \in (0,1].
	\end{equation}
Now the following proposition derives a necessary and sufficient condition for the optimal random variable   $S^{\star}$.
\begin{proposition}
\label{PropNecess}
$S^{\star}$ is the optimal random variable in $(0,1]$, if there exist $\lambda_1 \geq 0$ and $\lambda_2 \geq 0$, such that 
	\begin{IEEEeqnarray}{l}
	\nonumber
	\lambda_1\left(\frac1s-1-P\right)-\lambda_2\left(I_0(\sqrt{2}Bh_2\left(\sqrt{\frac1s-1}\right)-E_{\mathrm{req}}\right) \\+C-\log s+1
	+\int_0^{\infty}s\mathrm{e}^{-sy}\log p(y;F^{\star})\d y\geq 0,
	\end{IEEEeqnarray}
	for all $s \in (0,1]$, with equality if $ s \in \mathrm{Supp}(S^{\star}) $.
\end{proposition}

In the following, we  show that the equality in Proposition 1 can not be satisfied in a set that has an accumulation point\footnote{ A point $x$ is an accumulation point (limit point) of a set $\mathcal{A}$, if every open neighborhood of  $x$ contains at least one point from $\mathcal{A} \subset \mathcal{X}$ distinct from $x$.}, hence the support of $S^{\star}$ must be  discrete.  The discretness property of the optimal input distribution is given by Theorem \ref{TheoremDiscrete}.
\begin{theorem}
\label{TheoremDiscrete}
The optimal input distribution that achieves the capacity in \eqref{EqOptimization}, it is discrete with a finite number of mass points. 
\end{theorem}
\begin{remark}
 Note that the optimization problem in \eqref{EqOptimization} could be generalized to $K$ identical EH receivers; in this case,  the optimization problem is  written as
\begin{equation}
\label{EqOptimization2}
\begin{aligned}
& \underset{F \in \mathcal{F}}{\text{sup}}
& & I(F)=\int \int p(y|x)\log \frac{p(y|x)}{p(y; F)}\d y \d F(x), \\
& \text{subject to}
& &  \mathds{E}[X^2] \leq P, \\
& & & \mathcal{E}(X)\geq E_{k,\mathrm{req}}, \hspace{1ex} \forall \hspace{1ex} k \in \{1,2\ldots,K\}. 
\end{aligned}
\end{equation}
It has been shown in \cite{Rania-2020} that if $A <A_{s}$, then at most one EH constraint is active. Hence, the result of Theorem \ref{TheoremDiscrete} is applied for this case as well.  
\end{remark}
\begin{IEEEproof}
The proof is presented in Appendix \ref{ProofOfDiscrete}. \\
\end{IEEEproof}
\begin{remark}
\label{NumberofPoints}

 By using the fact that $F^{\star}$ is discrete with a finite number of mass points,  the problem in \eqref{EqOptimization} is reduced to the determination of the maximum for a function of a finite dimensional vector, where its  components are the mass points and their locations \cite{Smith-1971}. An arbitrary input distribution $F$ can be written as 
\begin{equation}
F(x)=\sum_{i=1}^N q_iu(x-x_i),    
\end{equation}
where $x \rightarrow u(x)$ is the unit step function. In practice, it would be interesting to impose a constraint on the number of mass points, i.e.,  $N <N_0$; since in practice, the constellation size of the transmitted signal is limited. In section \ref{SecNum}, we plot the information energy capacity region for different 
amplitude-shift keying (ASK) modulations. The original optimization problem in \eqref{EqOptVArOr} is written as 
	
	\begin{equation}
	\label{EqVec}
\begin{aligned}
& \underset{(x_1, x_2,..., x_n, q_1,q_2,..., q_N)}{\text{sup}}
& & I(x_1, x_2,..., x_N, q_1,q_2,..., q_N), \\
& \text{subject to}
& & \sum_{i=1}^N q_i=1,\\
& & & \sum_{i=1}^N q_i x_i^2 \leq P, \\
& & & \sum_{i=1}^N q_i\left[I_0\left(\sqrt{2}Bh_2x_i\right)\right]\geq E_{\mathrm{req}}, \\
& & &  N <N_0.
\end{aligned}
\end{equation}
Although the optimization problem in \eqref{EqOptimization} is convex over all input distribution functions, we can not claim that the problem in \eqref{EqVec} remains convex since it is paramatrized by the mass point probability and its locations \cite{Shemai-2001}. In practice, we discretize the interval  $[0,A]$, with sufficiently small step size $\Delta x$ \cite{Rania-2018} to obtain the mass points set. Then, we employ  a numerical solver such as CVX \cite{cvx} to solve \eqref{EqVec} by letting $\Delta x \rightarrow 0$.
\end{remark}
\subsection{Properties of the mass points}
 We give some insights on the behavior of the optimal input distribution $F^{\star}$ with respect to both the PP and the AP constraints. It has been shown that the optimal input distribution is discrete, therefore the CDF  is determined by the vectors $\bs{q}, \bs{x}$, and the number of the mass points $N$. Specifically, the location of the mass points is given by 
    \begin{equation}
			       \bs{x}=(x_1,\ldots,x_N),
			   \end{equation}
and the weights associated with the mass points,
	\begin{equation}
	    \bs{q}=(q_1,\ldots,q_N).
	\end{equation}
Without loss of generality, we assume that $ x_1 <x_2\ldots <x_n < A $. In addition, we assume that  $F^{\star}_{N}$ is the optimal input distribution and is  characterized by  the triplet $(\bs{q}^{\star},\bs{x}^{\star},N)$, which is the solution of the optimization problem in \eqref{EqOptimization}. The conditions which are satisfied by the optimal input distribution for some $\lambda_1 \geq 0$ and $\lambda_2 \geq 0$ are
	\begin{IEEEeqnarray}{lCl}
\nonumber
i(x;F^{\star}_N) &\leq& C\hspace{-0.5ex}+\hspace{-0.5ex}\lambda_1\left( x^2\hspace{-0.5ex}-\hspace{-0.5ex}P\right)\hspace{-0.5ex}-\hspace{-0.5ex}\lambda_2\left(\mathcal{E}(x)\hspace{-0.5ex}-\hspace{-0.5ex}E_{\mathrm{req}}\right), \hspace{.5ex} \text{for} \hspace{.5ex} x \in [0,A],\\
\nonumber
i(x_i^{\star};F^{\star}_N)&=&C+\lambda_1\left(x_i^{{\star}^2}\hspace{-0.5ex}-\hspace{-0.5ex}P\right)-\lambda_2\left(\mathcal{E}(x_i^{\star})\hspace{-0.5ex}-\hspace{-0.5ex}E_{\mathrm{req}}\right).\\
\end{IEEEeqnarray}
Denote by $g$ the following function i.e.,
\begin{IEEEeqnarray}{l}
g(w,F^{\star}_N)=i(w;F^{\star}_N) \hspace{-0.5ex}-\hspace{-0.5ex}\lambda_1\left( w^2\hspace{-0.5ex}-\hspace{-0.5ex}P\right)  \hspace{-0.5ex}+\hspace{-0.5ex}\lambda_2\left(\mathcal{E}(w)-E_{\mathrm{req}}\right).
\end{IEEEeqnarray}
It has been shown in \cite{Sharma-2010} that a point of increase (except $A$) is a local maximum for the function $g(x,F^{\star}_N)$, and hence 
\begin{equation}
\label{EqCondInput}
    \frac{\partial g(w,F^{\star}_N)}{\partial \omega}   \Bigg|_{w=x_i}=0.
\end{equation}
Unfortunately, a closed form solution for \eqref{EqCondInput} is an open problem even for the simpler case (AWGN channel). However, in the next remark, we are able to characterize a particular mass point for the optimal input distribution.
\begin{remark}
The input distribution has a necessary mass point at zero; by contradiction, we assume that  $0<x_1$, then it can be shown
\begin{equation}
    \frac{\partial g(w,F^{\star}_N)}{\partial \omega}   \Bigg|_{w=x_1} <0.
\end{equation}
Hence,  $x_1$ is not a point of increase for the function $f(x,F^{\star}_N)$ according to \eqref{EqCondInput}.
\end{remark}
In the following, we  study the behavior of the optimal input distribution at the transition point, where the binary distribution gives away to a ternary \cite{Sharma-2010}. Specifically,  we give a closed-form expression of the optimal input distribution on the transition point, where the binary distribution is not longer optimal. This is a critical point in SWIPT systems, since the binary distribution maximizes both information and energy transfer simultaneously and therefore there is not a trade-off between them \cite{9174172}. For simplicity, we assume that the AP constraint is active, for a low PP constraint, the optimal input distribution is binary and is characterized by $\bs{x}^{\star}$ and $\bs{q}$, i.e.,  
\begin{subequations}
\label{Eqbin}
\begin{IEEEeqnarray}{cCl}
\bs{x}^{\star}&=&(0,x_1),\\
\label{eqlocation}
\bs{q}&=&\left(1-\frac{P}{x_1^2},\frac{P}{x_1^2}\right).
\end{IEEEeqnarray}
\end{subequations}
Note that if 
\begin{equation}
    \frac{\partial I(Y,F^{\star}_N)}{\partial x_1}>0,
\end{equation}
then we have $x_1=A$. Let us assume that at the amplitude value $\bar{A}$, the binary distribution is not longer optimal. Thus at the amplitude value $\bar{A}+ \Delta \bar{A}$, a new mass point appears denoted by $x_2^{\star} \in [0,A]$, which satisfies   
\begin{equation}
    i(x_2^{\star};F_3^{\star}) =C+\lambda_1\left(x_2^{{\star}^2}-P\right)-\lambda_2\left(\mathcal{E}(x_2^{\star})-E_{\mathrm{req}}\right).
\end{equation}
Let $q$ the transition probability associated to the mass point $x_2^{\star}$, then the optimal input  distribution $F_3^{\star}$ is characterized  by 
\begin{IEEEeqnarray}{lCl}
\nonumber
q^{\star}&=&\Bigg(1-\frac{P}{(x_1+\Delta x_1)^2}-\left(1\hspace{-1mm}-\frac{x_2^{\star}}{(x_1\hspace{-1mm}+\hspace{-1mm}\Delta x_1)^2}\right)q, q,\\&&  \frac{P}{(x_1+\Delta x_1)^2}
- \frac{x_2^{{\star}^2}q}{(x_1+\Delta x_1)^2}\Bigg),\\
\bs{x}^{\star}&=&(0, x_2^{\star}, x_1+\Delta x_1).
\end{IEEEeqnarray}
\begin{remark}
\label{RemarkBinary}
Note at that for low PP constraints, there is not a trade-off between information and energy transfer, since the optimal input distribution is binary and hence it maximizes both information and energy transfer simultaneously. By increasing the PP constraint, we show that there  exist an amplitude value  $\bar{A}$, in which the binary distribution is not longer optimal and hence,  a trade-off between information and energy transfer is observed. The transition point $\bar{A}$ at the low PP regime,  satisfies the following  condition 
\begin{equation}
    I\left(F_3^{\star}(\bar{A}+\Delta \bar{A})\right) >   I\left(F_2^{\star}(\bar{A}+\Delta \bar{A})\right). 
\end{equation}
Hence by choosing  $\Delta$ small enough   ($\Delta \bar{A} \rightarrow 0$), we obtain a  sufficient condition for the transition point $\bar{A}$ \cite{4895581}. 
\end{remark}

\section{SWIPT for Time-Varying PP Constraints}\
\label{Sectime}
In this section, we  elaborate the structure of the capacity achieving input distribution for the general system model shown in Fig. 1.  For simplicity and  without loss of generality, we  assume that the PP constraint takes only two values $a_1$ and $a_2$ in a probabilistic way.  Note that in the case where the amplitude variation is available  at the transmitter, coding should be performed according the minimum amplitude. In this case, by applying Shannon's strategy \cite{Shann}, the extended input is reduced to  $\bs{X}=[X_1,X_2]$ that takes values into the set $\mathcal{X}= [0,A_1] \times [0,A_2]$, then the space of the admissible joint probability distribution function over $\mathcal{X}$ is given by  
			\begin{equation}
			    \mathcal{F}_{A_1,A_2}\defn \left\{ F : \int_{0}^{A_1} \int_{0}^{A_2} \d F(x_1,x_2)=1 \right\}.
			\end{equation}
The expression for  the equivalent channel  is given by 
\begin{IEEEeqnarray}{l}
\nonumber
 f(y|\hat{x}_1,\hat{x}_2)=p_1p(y|\Hat{x}_1)+p_2p(y|\hat{x}_2) \\
 \nonumber
=p_1\frac{1}{1+\hat{x_1}^2}\exp\left(-\frac{y}{1+\hat{x_1}^2}\right)+p_2\frac{1}{1+\hat{x_2}^2}\exp\left(-\frac{y}{1+\hat{x_2}^2}\right).\\
			\end{IEEEeqnarray}

In this case,  the mutual information that we aim to maximize is given by
\begin{equation}
    I(\bs{X};Y)=\int_{0}^{A_2} \int_{0}^{A_1}  \int f(y|x_1,x_2) \log \left(\frac{f(y|x_1,x_2)}{f(y;F)}\right) \d y \d F,
\end{equation}
where 
\begin{equation}
    f(y;F)=\int_{0}^{A_1} \int_{0}^{A_2} f(y|x_1,x_2) \d F(x_1,x_2), 
\end{equation}
with 
\begin{equation}
    f(y|x_1,x_2)=p_1p(y|x_1)+p_2p(y|x_2), 
\end{equation}
and the equivalent of the  mutual information density is given by 
\begin{equation}
    i(x_1,x_2;F)= \int f(y|x_1,x_2) \log \left(\frac{f(y|x_1,x_2)}{f(y;F)}\right) \d y.
\end{equation}
The original optimization problem in \eqref{EqOptVArOr}   is reduced to 
			\begin{equation}
			\label{EqOptVAr}
\begin{aligned}
& \underset{F \in {\mathcal{F}_{A_1,A_2}}}{\text{sup}}
& & I(F), \\
& \text{subject to}
& &  p_1\mathds{E}[X_1^2]+p_2\mathds{E}[X_2^2] \leq P, \\
& & &p_1\mathds{E}\left[\mathcal{E}(\hat{X}_1)\right]+p_2\mathds{E}\left[\mathcal{E}(\hat{X}_2)\right]\geq E_{\mathrm{req}},
\end{aligned}
\end{equation}
where
\begin{equation}
I(F)= \int \int \int f(y|x_1,x_2)\log \frac{f(y|x_1,x_2)}{f(y; F)}\d y \d F(x_1,x_2).
\end{equation}
In the following, we  prove that the optimal input distribution that solves the optimization problem in \eqref{EqOptVAr} has a support set of a finite cardinality. First, by using similar arguments as the static PP constrained case, we can establish  the uniqueness of the optimal input distribution; this follows from the following claims

 \begin{enumerate}
     \item $I(\bs{x};F)$ is a concave function and weakly differentiable in $\mathcal{F}_{A_1,A_2}$.
     \item $\mathcal{F}_{A_1,A_2}$ is a convex and compact space.
     \item The PP and AP constraints are linear.
     \item $i(x_1,x_2;F)$ is continuous and  analytical.
     \end{enumerate}

 By following a similar procedure as the static PP constraint case, the following theorem characterizes a necessary and sufficient condition for the optimal input distribution.
\begin{theorem}
\label{TheoVar}
 	$F^{\star}$ is the optimal input distribution, if there exist $\lambda_1 \geq 0$ and $\lambda_2 \geq 0$, such that 
	\begin{IEEEeqnarray}{lCl}
	\nonumber
	&&\lambda_1\left( p_1x_1^2+p_2x_2^2\hspace{-0.5ex}-\hspace{-0.5ex}P\right)\hspace{-0.5ex}-\hspace{-0.5ex}\lambda_2\big(p_1I_0(\sqrt{2}Bh_2\hat{x}_1)\hspace{-0.5ex}+\hspace{-0.5ex}p_2I_0(\sqrt{2}Bh_2\hat{x}_2)\\&&-E_{\mathrm{req}}\big)+C-\int p(y|x_1,x_2)\log \frac{p(y|x_1,x_2)}{p(y; F^{\star})}dy\geq 0,
	\end{IEEEeqnarray}
	for all $(x_1,x_2)$, with equality if $(x_1,x_2) \in \mathrm{Supp}(F^{\star})$.
\end{theorem}
By using the  invertible change of variables, 
\begin{equation}
    S=\left(\frac{1}{1+X_1^2},\frac{1}{1+X_2^2}\right), 
\end{equation}
a simpler necessary and sufficient condition on the optimal random variable  $S^{\star}=\left(\frac{1}{1+X_{1}^{{\star}^2}},\frac{1}{1+X_{2}^{{\star}^2}}\right)$  is derived in the following corollary.
\begin{corollary}
\label{CoroNec}
$S^{\star}$ is the optimal random variable, if and only if, there exist $\lambda_1 \geq 0$ and $\lambda_2 \geq 0$, such that 	
\begin{IEEEeqnarray}{lCl}
	\nonumber
	&&\lambda_1\left(p_1(\frac{1}{s_1}-1)+p_2(\frac{1}{s_2}-1)-a\right)-\lambda_2\bigg(p_1\mathcal{E}\left(\sqrt{\frac{1}{s_1}-1}\right)\\
	\nonumber
	&&+p_2\mathcal{E}\left(\sqrt{\frac{1}{s_2}-1}\right)-b\bigg)+C\\
	\nonumber
	&&+\int_0^{\infty} \sum_{i=1}^2 p_is_i\mathrm{e}^{-s_iy}\log \frac{\sum_{i=1}^2 p_is_i\mathrm{e}^{-s_iy}}{p(y)} dy \geq 0,
	\end{IEEEeqnarray}
	for all $(s_1,s_2)$, with equality if $(s_1,s_2) \in \mathrm{Supp}(S^{\star})$.
\end{corollary}
By using Corollary \ref{CoroNec}, we  establish that the optimal input distribution has a finite support.  
\begin{theorem}
\label{Theo:Time}
 The optimal input distribution $S^{\star}$ has a finite support.
\end{theorem}
\begin{IEEEproof}
The proof is presented in the Appendix \ref{ProofOfDiscreteTime}.
\end{IEEEproof}
\begin{remark}
Following the same procedure as \cite{Ozul},  we numerically compute the finite support of the capacity achieving distribution. Specifically, we fix the cardinality of the support as $|\mathrm{Supp}(F)|=2$, in this case, the optimal input distribution has two mass points characterized by $(x_1,x_2)=(0,0)$ and $(x_1,x_2)=(a_1,a_2)$. If this input  distribution  satisfies the necessary and sufficient condition in Theorem \ref{TheoVar}, then $F=F^{\star}$ otherwise we set the cardinality $|\mathrm{Supp}(F)|=3$ and the procedure is repeated. 
\end{remark}

\subsection{On-off energy arrivals}
We  consider the special case of on-off energy arrival, where at each channel use, either a constant amount  of energy  or a zero energy arrives at the transmitter \cite{Ozul} in a probabilistic way. The  alphabet of the amplitude random variable for this scenario is reduced to $\mathcal{A}=\{0,a_2\}$ where $p_2$ is  the probability that the transmitter receives energy. The optimization problem in \eqref{EqOptVArOn} is over the one dimensional distribution $F(0,x_2)$ and is given by 
			\begin{equation}
			\label{EqOptVArOn}
\begin{aligned}
& \underset{F_{X_2}\in {\Omega}}{\text{sup}}
& & I(F)= \int \int \int f(y|x_2)\log \frac{f(y|x_2)}{f(y; F_{X_2})}\d y \d F(x_2), \\
& \text{subject to}
& &  p_{2}\mathds{E}[X_2^2] \leq P, \\
& & &p_{2}\mathds{E}\left[\mathcal{E}\left(\hat{X}_2\right)\right]\geq E_{\mathrm{req}},
\end{aligned}
\end{equation}
where 
\begin{equation}
    f(y|x_2)=(1-p_{2})p(y)+p_{2}p(y|x_2).
\end{equation}

\begin{remark}
Similar to the static PP constraint case, for a low PP regime, i.e., $A \leq \bar{A}$, the optimal input distribution $F^{\star}$ is symmetric and binary with two mass points at $(0,A)$. Hence, there is not a trade-off between the information and the energy transfer. For the special case with $p_{2}=1$, the problem reduces to the static PP-constraint Rayleigh fading channel. 
\end{remark}

\begin{figure}[t]
\centering
\includegraphics[width=\linewidth]{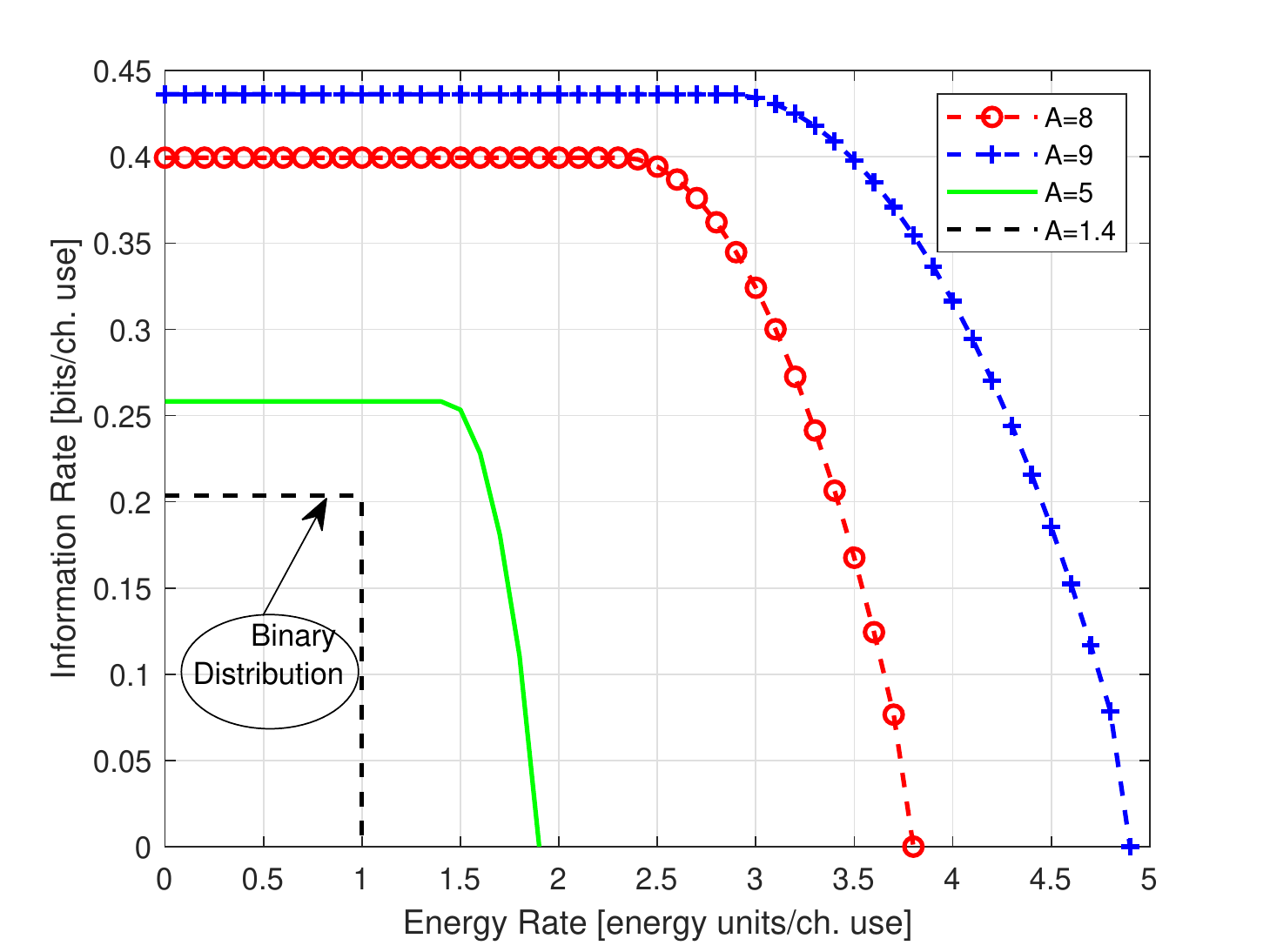}
\vspace{-0.3cm}
\caption{Information-energy capacity region for the static case with different PP constraints, $\beta=1$, $P=30$ dB, and $\sigma_1^2=\sigma_2^2=-80$ $\mathrm{dBm}$.} \label{fig2}
\end{figure}

\begin{figure}[t]
\centering
\includegraphics[width=\linewidth]{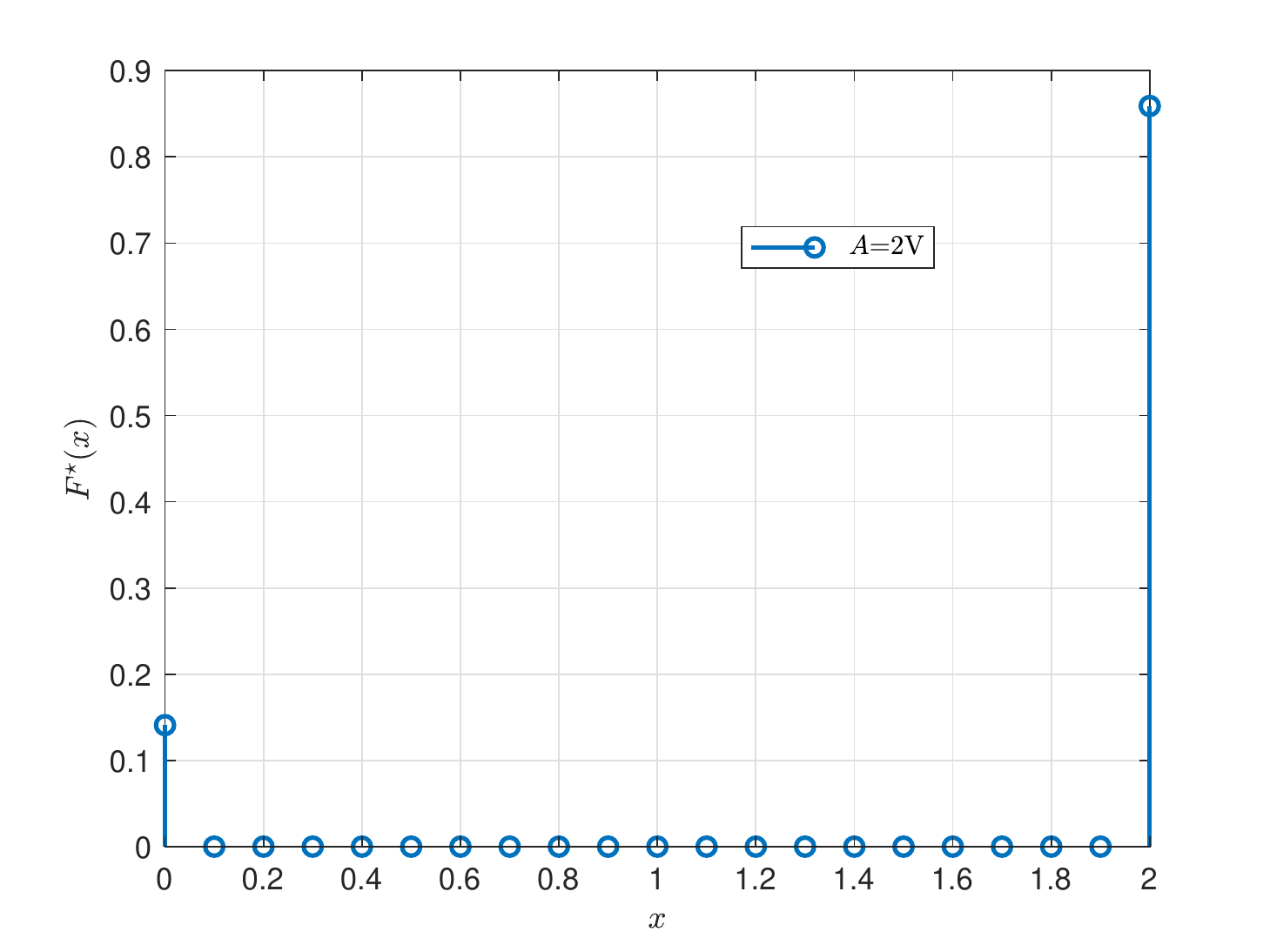}
\vspace{-0.3cm}
\caption{Optimal input distribution for the static case with a low PP constraint; $\beta=1$, $P=30$ dB, and $\sigma_1^2=\sigma_2^2=-80$ $\mathrm{dBm}$.} \label{fig6}
\end{figure}

\begin{figure}[t]
\centering
\includegraphics[width=\linewidth]{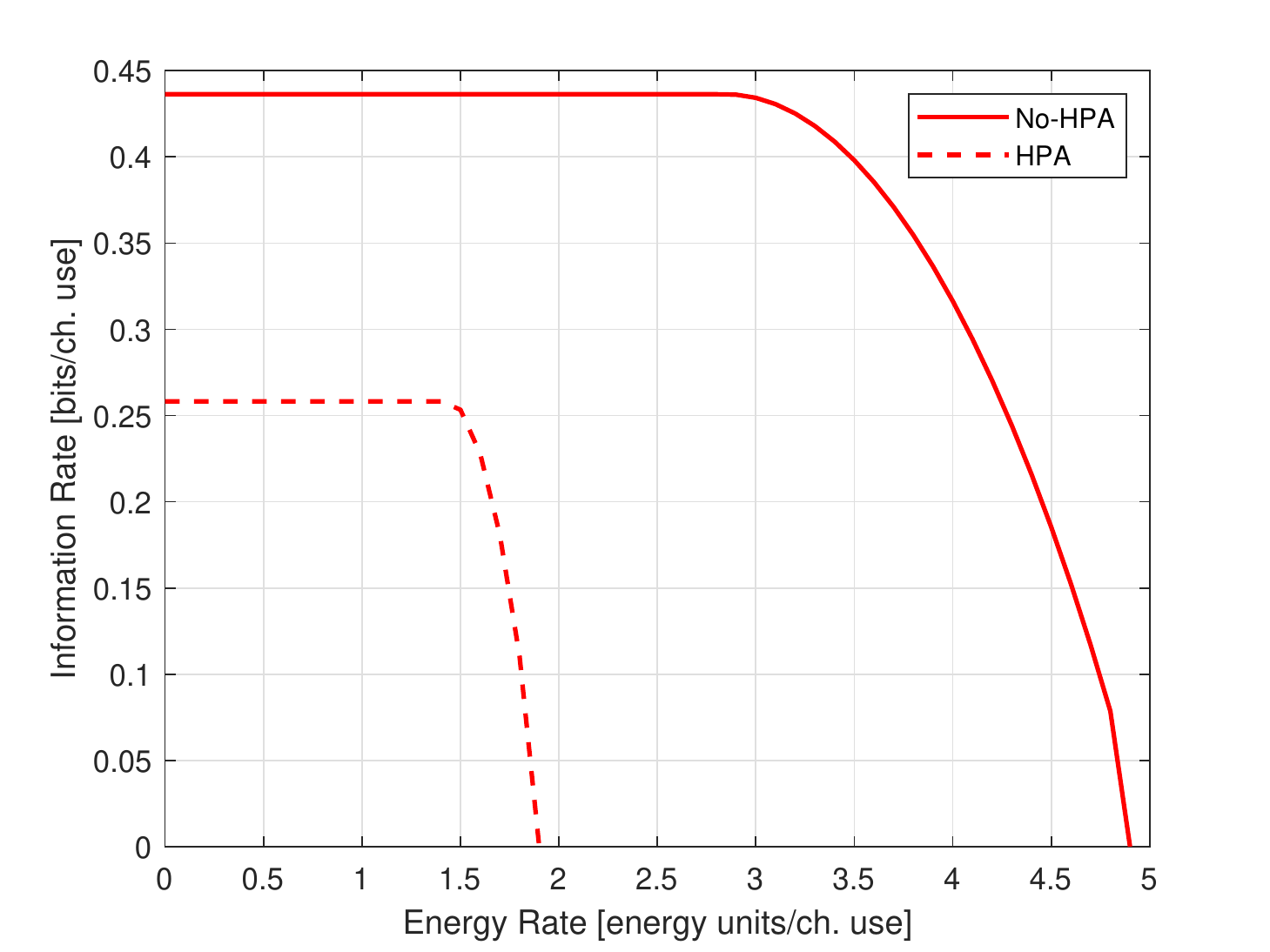}
\vspace{-0.3cm}
\caption{Effects of the HPA on the Information-energy capacity region; $A=5$, $\beta=1$, $B=0.5$, $P=30$ dB, and $\sigma_1^2=\sigma_2^2=-80$ $\mathrm{dBm}$.} \label{fig3}
\end{figure}

\begin{figure}[t]
\centering
\includegraphics[width=\linewidth]{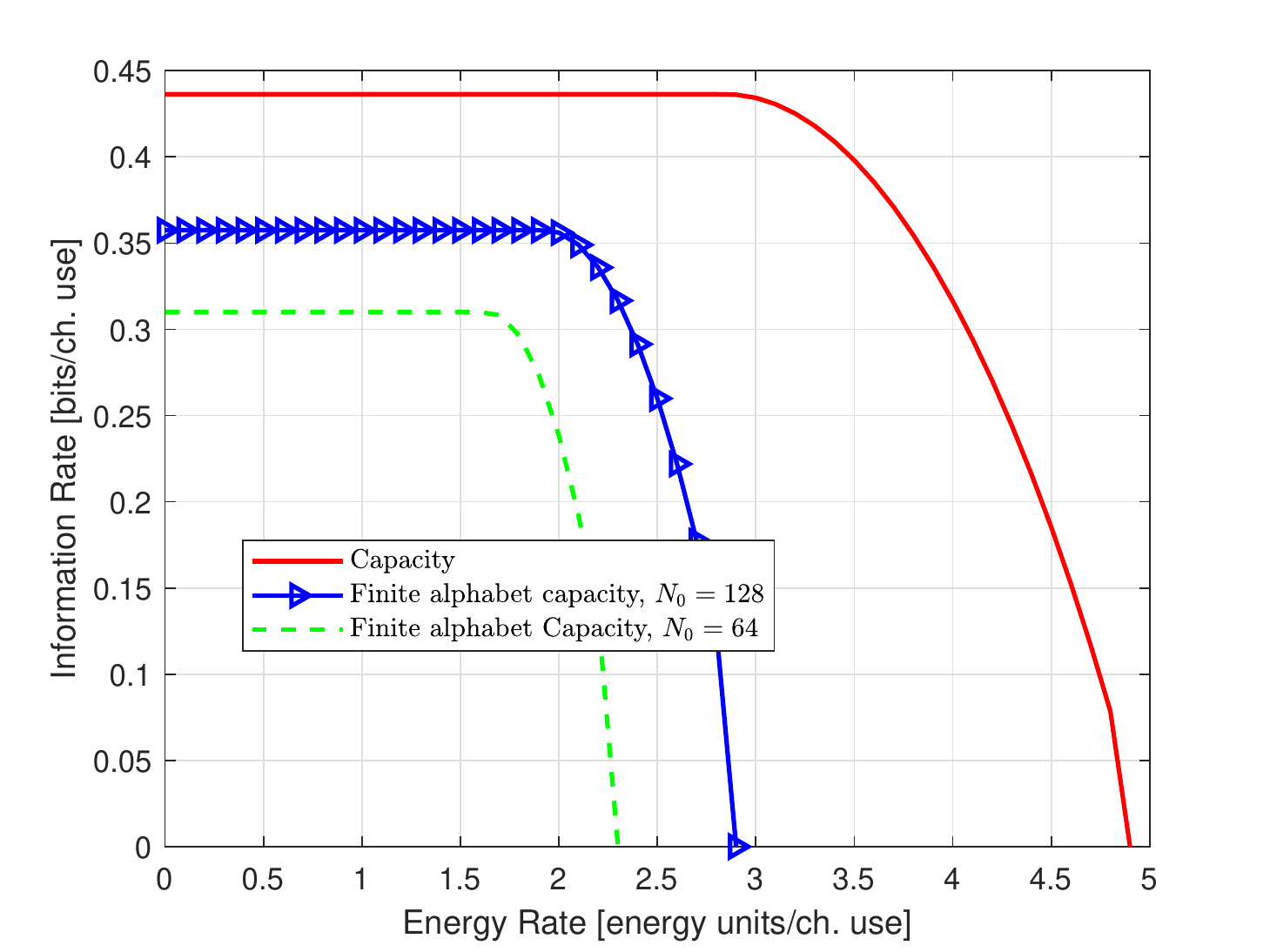}
\vspace{-0.3cm}
\caption{ Information energy capacity region for different finite alphabets $N_0$; $A=5$, $\beta=1$, $B=0.5$, $P=30$ $\mathrm{dB}$, and $\sigma_1^2=\sigma_2^2=-80$ $\mathrm{dBm}$.} \label{fig5}
\end{figure}

\begin{figure}[t]
\centering
\includegraphics[width=\linewidth]{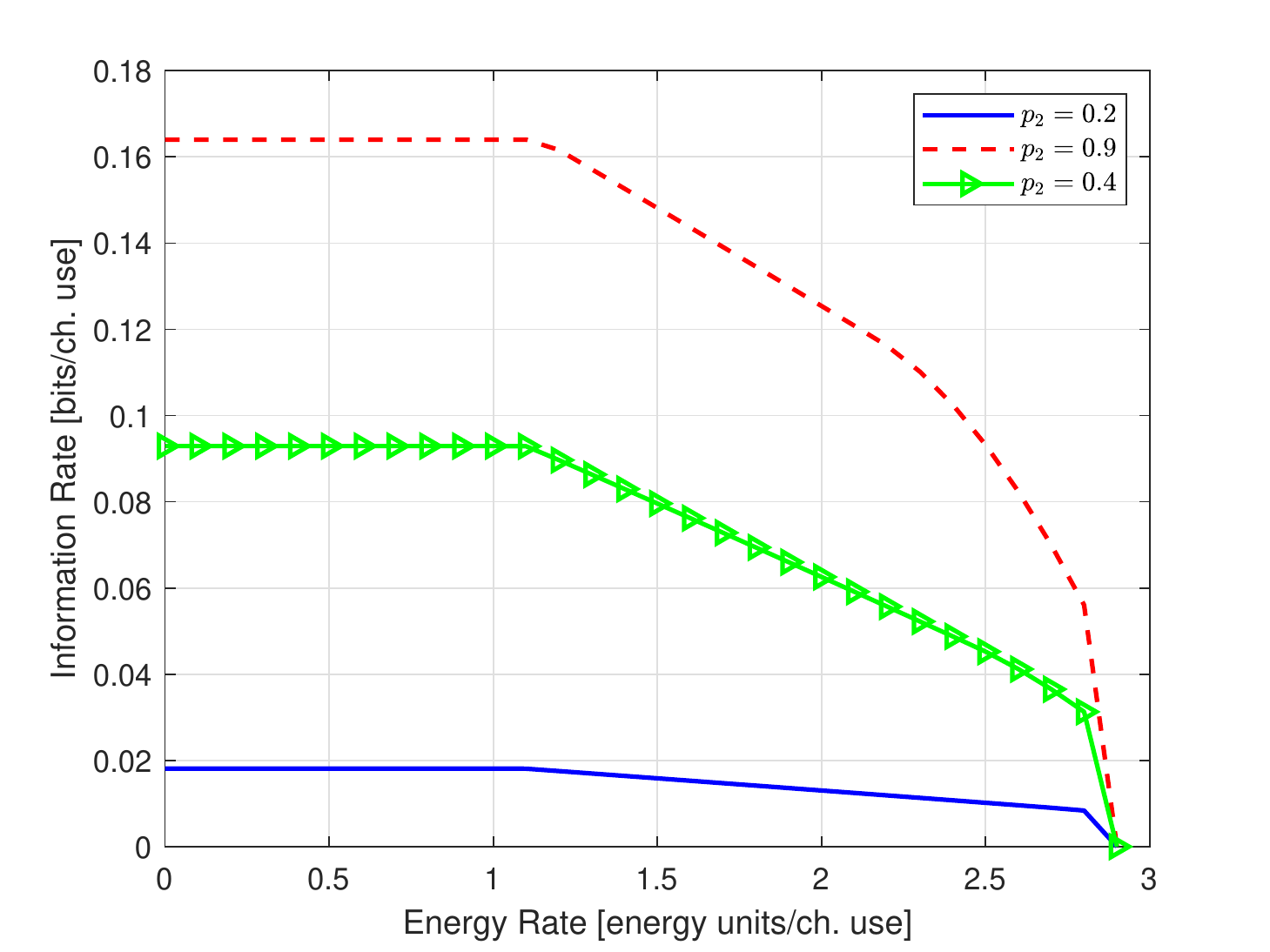}
\vspace{-0.3cm}
\caption{Effect of $p_{2}$ on the Information-energy capacity region for the time-varying case; $A=7$, $\beta=1$, $B=0.5$, $P=30$ dB, and $\sigma_1^2=\sigma_2^2=-80$ $\mathrm{dBm}$.} \label{fig4}
\end{figure}

\section{Numerical Results}
\label{SecNum}
 We numerically evaluate the information-energy capacity region by using a numerical solver such  CVX \cite{cvx}. For the scenario with a  static PP constraint, Fig. \ref{fig2} shows the information-energy capacity region for different PP constraints. The corresponding region is obtained by solving the optimization problem  in \eqref{EqOptimization}. A trade-off is observed between the information rate transmitted to the information receiver and the energy delivered to the EH receiver; this trade-off becomes evident since for  higher EH constraints, the transmitter selects a symbol with a higher amplitude, which on the other hand, it degrades the information transfer performance. Another interesting remark is that  for low  PP constraints, there is not a trade-off between the two objectives. The optimal input distribution for this regime is binary, hence it maximizes both information and energy transfer simultaneously (Remark \ref{RemarkBinary}). In Fig. \ref{fig6}, we examine the special case of small PP constraints i.e., $A=2V$; it can be seen the optimal input distribution is binary with two mass points \eqref{Eqbin}. This observation is also inline with Remark \ref{RemarkBinary}.

Fig. \ref{fig3} highlights the effect of the HPA non-linearity on  information-energy capacity region. There is a gap between  the two regions; this is mainly due to the negative effect of the HPA. Fig. \ref{fig5} plots the information energy capacity region for   ASK modulations, which is obtained by solving the optimization problem in Remark \ref{NumberofPoints} for different $N_0$ values. It can be seen that by increasing the alphabet size, a smaller gap is observed between the  rate achieved by the finite alphabet and the rate achieved by the optimal input distribution.
Finally, Fig. \ref{fig4} deals with the time-varying PP constraint case  under an on-off energy arrival, where the  state information is available causally for different values of $p_{2}$. We observe that an increase of $p_{2}$, enlarges the information energy capacity region.

\section{Concluding Remarks}
In this paper, we studied the fundamental limits of a SWIPT system over a Rayleigh- fading channel with time-varying constraints,  non-linear EH, and by taking into account the non-linearity imposed by the HPA. For the special case of static PP constraints, we proved that the input distribution that maximizes the information-energy capacity  region is unique, discrete, with a finite number of mass points. We have shown that the information energy capacity region increases by relaxing the PP constraint, while  the HPA significantly degrades the performance of both objectives. Also, we proposed a mathematical framework to study the capacity achieving distribution for low PP constraints, where there is not a trade-off between information and energy transfer. For the case with time-varying constraints, we have shown that the capacity achieving distribution has a finite support. Finally, we have  studied the optimal input distribution for the particular scenario  of on-off energy arrival. In our future work, we intend to concentrate on the fundamental limits of SWIPT for more general settings e.g., Rician fading channel.

%

\begin{appendices}

 \section{Proof of Theorem \ref{TheoremUnique}}
 \label{ProofOfTheoremunique}
Despite a great deal of similarity with the proof in \cite{Shemai-2001}, a sketch of the proof is presented for the  sake of completeness.   To prove the existence and the uniqueness of the optimal input distribution, it suffices to prove that the problem in \eqref{EqOptimization} is convex, i.e., we need to prove the following points 
\begin{enumerate}[label=(\roman*)]
	\item  The mutual information as function of the distribution $F$, i.e, $I: F\rightarrow I(F)$ is a weak continuous function.
	\item $I: F\rightarrow I(F)$ is strictly concave function.
	\item The set $\Omega$ is compact.
	\end{enumerate}
\emph{Continuity}: By using the fact that the weak topology on the distribution functions is metrizable,  we need to show that  for any sequence of probability distributions  $F^{(n)}$ that converges to  $F$,   $I(F^{(n)})$ converges to  $I(F)$. The mutual information for a specific distribution $F$ could be written as
\begin{equation}
    I(F)=h_Y(F)-h_{Y|X}(F),
\end{equation}
where
\begin{IEEEeqnarray}{cCl}
	   h_Y(F) &\defn& -\int_{-\infty}^{\infty}p(y;F) \log p(y;F) dy,\\
	   h_{Y|X}(F)&\defn& -\int_{-\infty}^{\infty}\int_{-A}^A p(y|x) \log p(y|x) \d F(x) \d y,
	\end{IEEEeqnarray}
	are  the entropy of the random variables $X$ and $X|Y$ respectively.
First, we need to show that the function $F \rightarrow h_Y(F)$ is weak continuous. Let $F^{(n)} \rightarrow F$, we need to show that $h_Y(F^{(n)}) \rightarrow h_Y(F)$ by establishing the following equality  
	    \begin{IEEEeqnarray}{lCl}
	    \label{Eq1}
	    \lim_{n}h_Y(F^{(n)})&=&-\lim_{n} \int p(y;F^{(n)}) \log p(y;F^{(n)}) dy\\
	    \label{Eq2}
	    &=& -\int \lim_{n} p(y;F^{(n)}) \log p(y;F^{(n)}) dy\\
	    \label{Eq3}
	    &=& - \int  p(y;F) \log p(y;F) dy\\
	    \label{Eq4}
	    &=&h_Y(F),
	    \end{IEEEeqnarray}
where \eqref{Eq1} and \eqref{Eq4} are from definition; \eqref{Eq2} follows from Lesbeque dominated convergence Theorem \cite{Feller-book-1971}; \eqref{Eq3} follows from the continuity of the function $x \rightarrow x\log x$. The proof of continuity of $F \rightarrow h_{Y|X}(F)$ follows the same arguments.\\
\emph{Concavity, compactness}: Note that $I(F)= h_Y(F)-h_{Y|F}(F)$ and since the function $P_Y \rightarrow h_Y(P_Y)$ is a strictly concave function, and by using the fact that $F \rightarrow P_Y(F)$ is a linear function, it holds that $F \rightarrow h_Y(F)$ is a strictly concave. In addition, the function $F \rightarrow h_{Y|X}(F)$ is a linear function, then $I: F\rightarrow I(F)$  is a strictly concave function.
The proof of the compactness of $\Omega$ is similar as \cite{Shemai-2001}.

 \section{Proof of Corollary  \ref{CoroLagrangien}}
 \label{ProofOfCoroLagrangien}
Define a  linear vector space $X$,  a normed space $Z$,  a convex subset of $X$ denoted by $\mathcal{F}$, and a positive cone  in $Z$  that contains an interior point.  Let $f$ be a real-valued concave functional on $\mathcal{F}$ and $g$ is a convex mapping from $\mathcal{F}$ to $Z$. Assume the existence of a point $F_1 \in \mathcal{F}$, for which $g(F_1) <0$ (Slater's condition). Let 
 \begin{equation}
 \label{EqOpt}
 C=\sup_{F \in \mathcal{F} \\ g(F) \leq 0} f(F),
 \end{equation}
and assume $C$ is finite. Then  by the Lagrangian theorem \cite{Smith-1971},  there is an element $z_0^{\star} \in Z$ such that 
\begin{equation}
\label{EqScal}
    C=\sup_{F \in \mathcal{F}} \{f(F)-\langle {g(F),z_0^{\star}}\rangle\},
\end{equation}
where $\langle\cdot,\cdot\rangle$ denotes the scalar product.
Note that if  $F^{\star}$ is the solution of the optimization problem in \eqref{EqOpt}, then $C$ also is achieved by $F^{\star}$ in \eqref{EqScal} and 
\begin{equation}
    \langle {g(F^{\star}),z_0^{\star}} \rangle=0.
\end{equation}
Now we need to verify that the Slater's condition holds, i.e there exist an interior point $F \in \Omega$ such that all the constraints hold with a strict inequality, i.e $g_i(F)<0, i=1,2$. Let $x_1$ satisfies $|x_1|<0$ and $\mathcal{E}(\hat{x}_1) >E_{\mathrm{req}}$ and consider $F_1$ the step function at $x_1$, then the following holds 
\begin{IEEEeqnarray}{cCl}
  g_1(F_1)&=&x_1^2-P <0,\\
  g_2(F_1)&=&-\mathcal{E}(\hat{x}_1) +E_{{req}}<0.
\end{IEEEeqnarray}
Hence the conditions of the Lagrangian theorem are satisfied, which completes the proof. 

 \section{Proof of Theorem \ref{TheoremNecessary}}
 \label{ProofOfNecessarry}
 The proof relies on the basic optimization Theorem \cite{Smith-1971}, that determine a necessary and sufficient condition for the optimal solution of the following optimization problem. Let  $f$ be a continuous,  and  weakly-differentiable,  and strictly convex map from a compact and convex space  $\Omega$ to $\mathds{R}$. Define also
	       \begin{equation}
	           C\defn \underset{x \in \Omega}{\text{sup}}f(x),
	       \end{equation}
	  then  the following claims hold, 
	       \begin{enumerate}
	           \item $C=\text{max}f(x)=f(x_0)$ for some unique $x_0 \in \Omega$.
	           \item A necessary and sufficient condition for $f(x_0)=C$ is $f^{'}_{x_0}(x) \leq 0$.
	            \end{enumerate}
	            
Since $\Omega$ is a convex, and 
$J(F)\defn I(F)-\lambda_1g_1(F)-\lambda_2g_2(F)$  is weakly differentiable, then by applying the optimization theorem,  $J^{'}_{F^{\star}}(F) \leq 0$ is a necessary and sufficient condition for $J(F)$ to achieve its maximum on $F^{\star}$, where $J^{'}_{F^{\star}}(F)$ is defined as follows
\begin{equation}
    J^{'}_{F^{\star}}(F)=\lim_{\theta \rightarrow 0} \frac{J\left((1-\theta)F^{\star}+\theta F\right)-J(F^{\star})}{\theta}.
\end{equation}
It remains to prove that $F \rightarrow J(F)$ is weakly differentiable and  determine its first derivative.
Define 
\begin{equation}
    F_{\theta}=(1-\theta)F^{\star}+\theta F.
\end{equation}
Then 
\begin{IEEEeqnarray}{l}
 \nonumber
    I( F_{\theta})-I(F^{\star})=\int \int p(y|x) \log \frac{p(y;F^{\star})}{p(y;F_{\theta}} \d y \d F^{\star}(x)\\
    +\theta \int \int i(x;F_{\theta}) \d F(x)-\theta \int \int i(x;F_{\theta}) \d F^{\star}(x).
\end{IEEEeqnarray}
Since 	 
\begin{equation}
	        p(y;(1-\theta)F^{\star}+\theta F)=(1-\theta)p(y;F^{\star})+\theta p(y;F),
	    \end{equation}
	    the following expression holds 
	    \begin{IEEEeqnarray}{lCl}
	    \nonumber
	       I_{F^{\star}}^{'}(F)&=&\lim_{\theta \rightarrow }\frac{I(F_{\theta})-I(F^{\star})}{\theta}=\int i(x;F^{\star}) \d F(x)-I(F^{\star}).\\
	    \end{IEEEeqnarray}
	    It has been shown that for the linear constraints, 
	    \begin{equation}
	        g^{'}_{i,F^{\star}}(F)=g_i(F)-g_i(F^{\star}),
	    \end{equation}
and from the complementary slackness conditions,
\begin{equation}
    g_i(F^{\star})=0, \hspace{2ex} i \in\{1,2\}.
\end{equation}
Hence  the condition $J^{'}_{F^{\star}}(F) \leq 0$ implies
\begin{equation}
    \int i(x;F^{\star}) \hspace{-0.5ex}- \hspace{-0.5ex} C \hspace{-0.5ex}- \hspace{-0.5ex}\lambda_1g_1(F) \hspace{-0.5ex}- \hspace{-0.5ex}\lambda_2g_2(F) \d F(x) \leq C-\lambda_1P+\lambda_2E_{\mathrm{req}},
\end{equation}
 which completes the proof.

\section{Proof of Corollary \ref{CoroNecessary}}
 \label{ProofOfCoroNecess}

Let $E_0$ be the points of increase of a distribution function $F^{\star}$  and define 
\begin{IEEEeqnarray}{cCl}
  A_1(x)&=&x^2, \\
  A_2(x)&=&-\mathcal{E}(\hat{x}) ,\\
  a_1&=&P, \\
  a_2&=&-E_{\mathrm{req}}.
\end{IEEEeqnarray}
Thus we can write the inequality in Theorem \ref{TheoremNecessary} as  following
\begin{equation}
\label{EqCond1}
\int \left(i(x;F^{\star})-\sum_{i=1}^2 \lambda_iA_i(x)\right) \d F(x) \leq C-\sum_{i=1}^2 a_i^2.
\end{equation}
Now, we need to prove that \eqref{EqCond1} is satisfied if and only if 
\begin{equation}
\label{EqCond2}
    i(x;F^{\star}) \leq C+\sum_{i=1}^2 \lambda_i(A_i(x)-a_i), \hspace{2ex} \text{for } x \in [0,A],
\end{equation}
and 
\begin{equation}
\label{EqCond3}
    i(x;F^{\star}) = C+\sum_{i=1}^2 \lambda_i(A_i(x)-a_i), \hspace{2ex} \text{for } x \in E_0.
\end{equation}
Note that if \eqref{EqCond2} and \eqref{EqCond3} hold  then \eqref{EqCond1} is satisfied immediately. The second part of the proof is to prove (by contradiction)  that if \eqref{EqCond1} holds then \eqref{EqCond2} and \eqref{EqCond3} are satisfied. First, we  assume that if the inequality in \eqref{EqCond2} is false then there exist $\tilde{x}$, such that 
\begin{equation}
\label{EqContra}
    i(\tilde{x};F^{\star}) > C+\sum_{i=1}^2 \lambda_i(A_i(\tilde{x})-a_i), \hspace{2ex} \text{for all } x \in E_0.
\end{equation}
Since \eqref{EqContra} holds $\forall \hspace{1ex} F \in \Omega$, then by choosing a particular distribution  as the step function at $\tilde{x}$, the following holds 
\begin{equation}
    i(\tilde{x};F^{\star}) > C+\sum_{i=1}^2 \lambda_i(A_i(\tilde{x})-a_i),
\end{equation}
which is a contradiction to the inequality in \eqref{EqCond1}. Now assume \eqref{EqCond2} holds but not \eqref{EqCond3}, then there exist $\tilde{x} \in  E_0$ such that 
\begin{equation}
\label{EqCond4}
    i(\tilde{x};F^{\star}) > C+\sum_{i=1}^2 \lambda_i(A_i(\tilde{x})-a_i).
\end{equation}
By using the continuity of the functions in \eqref{EqCond4}, then there exist a set $E^{'}$ (neighborhood of $\tilde{x}$) with a non zero measure, i.e, $\int_{E^{'}} \d F^{\star}(x)= \delta >0$, such that  \eqref{EqCond4} holds. Hence,
\begin{IEEEeqnarray}{lCl}
\nonumber
  C-\sum_{i=1}^2 \lambda_i a_i &=&I(F^{\star})-\sum_{i=1}^2\lambda_i \int A_i(x) \d F^{\star}(x)\\
  \nonumber
  &=& \int \left(i(x;F^{\star})-\sum_{i=1}^2 \lambda_iA_i(x)\right) \d F^{\star}(x)\\
  \nonumber
  &=&\int_{E^{'}} \left(i(x;F^{\star})-\sum_{i=1}^2 \lambda_iA_i(x)\right) \d F^{\star}(x)\\
  \nonumber
 &&+\int_{E-E^{'}}\left(i(x;F^{\star})-\sum_{i=1}^2 \lambda_iA_i(x)\right) \d F^{\star}(x)\\
 \nonumber
  &<&\delta\left(C-\sum_{i=1}^2\lambda_i a_i\right)\hspace{-0.5ex}+\hspace{-0.5ex}(1-\delta)\left(C-\sum_{i=1}^2\lambda_i a_i\right)\\
  &<& C-\sum_{i=1}^2\lambda_i a_i,
\end{IEEEeqnarray}
which is a contradiction. This completes the proof.
 \section{Proof of Theorem \ref{TheoremDiscrete}}
 \label{ProofOfDiscrete}

 Assuming that $S^{\star}$ is not discrete, and $\mathrm{Supp}(S^{\star}) \subset [0,A]$ then $\mathrm{Supp}(S^{\star})$ has a limit point by the Bolzano-Weierstrass theorem \cite{Feller-book-1971}. Denote by $h: z \rightarrow h(z)$ the following function
	\begin{IEEEeqnarray}{lCl}
	\nonumber
	h(z)=\lambda_1\left(\frac1z-1-a\right)-\lambda_2\left(I_0(\sqrt{2}Bh_2\left(\sqrt{\frac1z-1}\right)-b\right)\\
	\nonumber
	+C-\log z+1	+\int_0^{\infty}z\mathrm{e}^{-zy}\log p(y)dy, \hspace{1ex} z \in \mathcal{D},
	\end{IEEEeqnarray}
	with $\mathcal{D}$ defined by $\Re (z) >0$.   By extending the necessary and sufficient conditions of Proposition \ref{PropNecess} to the complex domain, we have 
	\begin{equation}
	    h(z)=0, \hspace{1ex} z \in \mathrm{Supp}(S^{\star}).
	\end{equation}
	
	 Recall that the support of $S^{\star}$ has an accumulation point and the function $h(z)$ is analytic over the domain  $\mathcal{D}$, hence by applying the identity theorem \cite{Smith-1971}
	 	\begin{equation}
	 	\label{EqAnaly}
	    h(z)=0, \hspace{1ex} z \in \mathcal{D}.
	\end{equation}
	By using \eqref{EqAnaly}, then  we have
\begin{IEEEeqnarray}{l}
	\nonumber
	\int_0^{\infty}s\mathrm{e}^{-sy}\log p(y)dy=-\frac{1}{s}\bigg[\lambda_1\left(\frac1s-1-a\right)\\
	\nonumber
	-\lambda_2\left(I_0(\sqrt{2}Bh_2\left(\sqrt{\frac1s-1}\right)-b\right)	+C-\log s+1\bigg], \\
	\hspace{2ex} \forall \hspace{1ex} s \in ]0,1]
	\label{Eqleft}
.
	\end{IEEEeqnarray}
 The  left hand side in \eqref{Eqleft} is the unilateral Laplace transform of the function $\log p(y)$, while the right-hand side  (without the Bessel function \cite{Bowman-1958}) can be recognized as the Laplace transform of 
	\begin{equation}
		-\lambda_1y+\left[\lambda_1(1+a)-C-1-C_E\right]-\log y,
	\end{equation}
	where $C_E$ is Euler's constant.
The modified Bessel function is given by 	
\begin{IEEEeqnarray}{l}
\nonumber
  I_0\left(\sqrt{2}Bh_2(\sqrt{\frac1s-1})\right)=\sum_{n=0}^n a_n \left(\frac{1}{s}-1\right)^n\\
  \nonumber
  =\sum_{n=0}^{\infty} a_n \sum_{k=0}^{n}\binom{n}{k}\frac{1}{s^k}(-1)^{n-k},
\end{IEEEeqnarray}
with $a_n=\frac{(Bh_2/\sqrt{2})^{2n}}{n!^2}$.
	 By using the fact that
	\begin{equation}
    	\mathcal{L}^{-1}\left(\frac{1}{s^k}\right)=\frac{y^{n-1}}{n!},
    \end{equation}
and by taking into account the uniqueness of the Laplace transform for continuous functions with bounded variation, the following expression holds 

	    	\begin{IEEEeqnarray}{l}
	    	\nonumber
	    	    p(y)=K\frac{\exp(-\lambda_1y)}{y}\\
	    	    \nonumber
	    	    \times \exp\left(\lambda_2\sum_{n=0}^{\infty} a_n \sum_{k=0}^{n}\binom{n}{k}\frac{y^k}{(k+1)!}(-1)^{n-k}\right).
	    		\end{IEEEeqnarray}
For every $\lambda_1 > 0$ and $\lambda_2 > 0$, we have
\begin{equation}
    \int_0^{\infty} p(y)dy >\infty,
\end{equation}
hence, $p(y)$  cannot be a probability distribution, and $\mathrm{Supp}(S^{\star})$ cannot have an accumulation point; which means that the optimal input distribution is discrete. 

 \section{Proof of Theorem \ref{Theo:Time}}
 \label{ProofOfDiscreteTime}
 Denote by $K:(s_1,s_2) \rightarrow K(s_1,s_2)$  the following function as 
     \begin{IEEEeqnarray}{l}
	\nonumber
	K(s_1,s_2)=\lambda_1\left(p_1(\frac{1}{s_1}-1)+p_2(\frac{1}{s_2}-1)-a\right)\\
	\nonumber
	-\lambda_2\bigg(p_1I_0(\sqrt{2}Bh_2\left(\sqrt{\frac{1}{s_1}\hspace{-0.5ex}-\hspace{-0.5ex}1}\right)
	\hspace{-0.5ex}+\hspace{-0.5ex}p_2I_0(\sqrt{2}Bh_2\left(\sqrt{\frac{1}{s_2}\hspace{-0.5ex}-\hspace{-0.5ex}1}\right)\hspace{-0.5ex}-\hspace{-0.5ex}b\bigg)\\
	+C+\int_0^{\infty} \sum_{i=1}^2 p_is_i\mathrm{e}^{-s_iy}\log \frac{\sum_{i=1}^2 p_is_i\mathrm{e}^{-s_iy}}{p(y)} dy. 
	\end{IEEEeqnarray}
 
Assuming that $\mathrm{Supp}(S^{\star})$ includes infinitely many elements in $\mathds{R}^2$, then  $\mathrm{Supp}(S^{\star})$ has an accumulation point, and since $K(s_1,s_2)$ is an  analytical function, then by applying the identity theorem \cite{Feller-book-1971}, the following expression holds
\begin{equation}
    K(s_1,s_2)=0, \hspace{2ex} \text{for all} \hspace{1ex} (s_1,s_2) \in (0,1]^2.
\end{equation}
By taking $s_1=s_2=s$, then 
\begin{equation}
    K(s_1=s,s_2=s)=h(s),
\end{equation}
which causes a contradiction by following the same procedure with Appendix \ref{ProofOfDiscrete}.
\end{appendices}

\balance
\bibliographystyle{IEEEtran}
\bibliography{journalpaper.bib}
\balance

\end{document}